# Size-dependent spin-reorientation transition in $Nd_2Fe_{14}B$ nanoparticles

*Chuan-bing Rong, Narayan Poudyal, and J. Ping Liu*

*Department of Physics, University of Texas at Arlington, Arlington, TX 76019*

## Abstract

$Nd_2Fe_{14}B$ magnetic nanoparticles have been successfully produced using a surfactant-assisted ball milling technique. The nanoparticles with different size about 6, 20 and 300 nm were obtained by a size-selection process. Spin-reorientation transition temperature of the NdFeB nanoparticles was then determined by measuring the temperature dependence of DC and AC magnetic susceptibility. It was found that the spin-reorientation transition temperature ($T_{sr}$) of the nanoparticles is strongly size dependent, i.e., $T_{sr}$ of the 300 nm particles is lower than that of raw materials and a significant decrease was observed in the 20 nm particles.

## 1. Introduction

Magnetic nanoparticles with high magnetocrystalline anisotropy, such as FePt, $SmCo_5$ and $Nd_2Fe_{14}B$, have drawn great attention because of their potential applications in high density magnetic recording and high energy-product permanent magnets.[1-2] It is of fundamental and technical interests to understand the correlation between particle size and properties of these high-anisotropy nanoparticles. The size dependence of Curie temperature has been recently observed in many magnetic materials, such as Gd thin films,[3] $MnFe_2O_4$,[4] and FePt nanoparticles.[5] However, the size effect on the magnetic properties of rare-earth containing magnetic nanoparticles has not been studied yet due to the difficulty in preparing the nanoparticles, which are very unstable in ambient condition. Our recently developed surfactant-assisted ball milling technique has been proved to be an effective way to prepare rare-earth transition-metal nanoparticles, including $SmCo_x$ (x=3.5-10) and $Nd_2Fe_{14}B$,[6-9] which makes it possible to study the size-dependent magnetic properties of the important magnetic compounds. Spin reorientation is a temperature-induced magnetic phase transition observed in $R_2Fe_{14}B$ (R=rare earth) compounds. The spin-reorientation temperature ($T_{sr}$) is characterized by a temperature below which the easy magnetization direction deviates away from the *c*-axis. The spin-reorientation temperature corresponds to the temperature at which the anisotropy constant $K_1$=0. $Nd_2Fe_{14}B$ is a typical compound with a spin reorientation temperature about 135 K.[10-13] It was reported by Kou at el. that $T_{sr}$ of nanocrystalline $Nd_2Fe_{14}B$ ribbons with grain size about 20 nm can be reduced to 117 K which is lower than that of the micro- and single-crystalline magnets.[14] However, no spin-reorientation temperature of magnetic nanoparticles have been reported so far. In this work, we reported the size effect on the spin-reorientation transition temperature of $Nd_2Fe_{14}B$ nanoparticles. It is interesting to find that $T_{sr}$ of the nanoparticles can be further reduced compared with the nanocrystalline materials and far lower than that of micro-crystalline materials.

## 2. Experimental procedure

The starting materials were firstly mixed with organic solvent heptane and surfactants (90% purity oleic acid and 98% purity oleyl amine) in a glove box. The mixture was ground in a milling vial using a SPEX 8000M high energy ball milling machine. There are two types of raw materials in this work: One is arc-melted $Nd_2Fe_{14}B$ ingots from Electron Energy Corp (EEC ingots) and the other is MQP-C ribbons from Magnequench inc. The resultant slurry was then dispersed into heptane solvent for size selection which was controlled by the time of the settle-down process.[6] Samples for magnetic characterization were prepared by embedding the nanoparticles in epoxy inside a glove box. The magnetic properties were measured by a SQUID magnetometer and a physical properties measurement system (PPMS) from 5K to room temperature. Structural and morphological characterizations were made using x-ray diffraction (XRD), small angle x-ray scattering (SAXS), transmission electron microscope (TEM), scanning electron microscope (SEM), and energy dispersive x-ray (EDX) analysis.

## 3. Results and discussion

Figures 1 (a) and (b) show the TEM images of the floating-at-top and 3h settle-down $Nd_2Fe_{14}B$ nanoparticles. The preparation details of the floating-at-top, 3h settle-down and the later mentioned slurry particles can be found in Ref. [9]. The particle size of the floating-at-top and the 3h settle-down particles are determined to be about 7 and 20 nm, respectively. The SEM image of the slurry particles is given in Fig. 1(c) for comparison. It can be seen that the particle size of the slurry is about 300-500 nm, which is much larger than that of the floating-at-top and 3h-settle-down particles.

The homogeneity of the nanoparticles is very important in this work to study the size-dependent magnetic properties since a small amount of impurity with micro-crystalline particles will change the accuracy of the results. SAXS analysis can further confirm particle size and size distribution of nanoparticles of whole samples instead of only local information. Figure 2 gives the SAXS curves and the fitted size distribution of the floating-at-top and 3h-settle-down particles. The size distributions of the particles were obtained by analyzing the scattering profile, as shown in the inset of Fig. 2. The average diameters were estimated to be 6.5 and 20 nm for the floating-at-top and 3h-settle-down particles, respectively, which are consistent with the microscopy analysis. It should be noted that size distribution of the 3h-settle-down particles (50%) is much larger than that of the floating-at-top particles (20%), which also agrees with the TEM analysis.

Figure 3 shows the XRD patterns of the $Nd_2Fe_{14}B$ nanoparticles with different particle size. The XRD pattern of the raw material is also given for comparison. The grain size of the slurry particles can be calculated according to the broadened diffraction peaks. For the 3h settle-down particles, diffraction peaks are still observed which means the nanoparticles are still crystallized. However, there is only one broad peak for the floating-at-top nanoparticles, which means that these small particles are amorphous. We will not study the magnetic properties of the floating-at-top particles in this paper since they are amorphous. It should be mentioned that the 3h settle-down particles are mixed with epoxy to prevent oxidation during XRD measurement, which dramatically decreases the X-ray diffraction intensity and increase the background. To confirm the structure, Figure 3(b) shows the XRD pattern of the 3h settle-down nanoparticles after exposing them in air for 24 hours without epoxy, which has made the nanoparticles completely oxidized. As one can see, the XRD pattern of 3h settle-down particles is very clear with strong $Nd_2O_3$ peaks. The estimated grain size is about 20 nm which agrees with the TEM analysis. This is an indirect evidence that the 3h settle-down particles are crystalline.

To determine the spin-reorientation transition temperature of NdFeB nanoparticles, we performed both DC and AC magnetic measurements. For the DC method, we measured the M-T curves with an applied magnetic field of 1 kOe and $T_{sr}$ was determined from the dM/dT curves. For the AC method, the temperature dependence of the ac susceptibility was measured from 5 to 350 K with a frequency from 10 to 10 kHz and an ac field up to 10 Oe. The real component ($\chi'$) and the imaginary component ($\chi''$) then can be used to determine $T_{sr}$. Figure 4 shows the DC and AC measurements of the products made from MQP-C ribbons as an example. Table 1 summarizes the $T_{sr}$ of the $Nd_2Fe_{14}B$ nanoparticles. It is evident that the spin-reorientation transition temperature $T_{sr}$ decreases with decreasing particle size. For example, $T_{sr}$ is about 135±3, 118±5, 103±5 K for the raw material, slurry and 3h settle down particles made from EEC ingot, respectively. Similar to that reported in Ref. 6, the $T_{sr}$ of the nanocrystalline slurry with submicron particle size is lower than that of raw materials, which could be attributed to the change of anisotropy constant $K_1$ due to strong intergrain exchange coupling.[14] More interestingly, it is for the first time to observe that the $T_{sr}$ of nanoparticles is even lower than that of nanocrystalline magnets. The mechanism of the particle size dependence of spin-reorientation temperature is to be further understood. It may be reasonable to correlate the size effect of spin-reorientation temperature to the change in magnetocrystalline anisotropy due to the surface modification and induced strain during ball milling. It was noted that the spin-reorientation behavior disappeared for the floating-at-top nanoparticles which are amorphous. The size effect on the spin-reorientation temperature has also been confirmed in the $Nd_2Fe_{14}B$ nanoparticles prepared by MQP-C ribbons. Different from the particles made by ingot, the $T_{sr}$ of slurry (113±5 K) is only slightly lower than that of raw materials (120±3 K), since the ribbons are already nanocrystalline while the surface effect is very small for the slurry particles. Further reduce the particle size to 20 nm significantly decreases the $T_{sr}$ down to 88±5 K, which confirms that $T_{sr}$ of nanoparticles is lower than that of nanocrystalline magnets and far lower than that of microcrystalline magnets.

## 4. Conclusions

In summary, the $Nd_2Fe_{14}B$ nanoparticles with different size have been prepared by surfactant-assisted ball milling. The structure of the nanoparticles has been determined as the tetragonal phase when particle size is equal or larger than 20 nm. The spin-reorientation transition temperature decreases remarkably from 135 K for the microcrystalline bulks to ~120 K for the nanocrystalline submicron particles. It can be further decreased to 90-100 K by decreasing particle size to 20 nm. The size effect on spin-reorientation transition temperature may be related to the decreased magnetocrystalline anisotropy.

## Acknowledgments

This work was supported by DARPA/ARO under grant W911NF-08-1-0249. This work is also supported by Center of Nanostructured Materials and Characterization Center for Materials and Biology at the University of Texas at Arlington. The authors would like to express appreciations to Dr Jinfang Liu of Electron Energy Corp. for providing the NdFeB alloy powders and Dr. Jim Herchenroeder Dr. Zhongmin Chen of Magnequench Co. for providing the MQP-C ribbons. The authors thank Dr. R. Skomski for a useful discussion.

**Reference**

1. R. Skomski and J. M. D. Coey, Phys. Rev. B, 48, 15812 (1993).

2. C.B. Rong, H.W. Zhang, R.J. Chen, S.L. He, B.G. Shen, J. Magn. Magn.Mater. 302, 126 (2006).

3. M. Farle, K. Baberschke, U. Stetter, A. Aspelmeier and F. Gerhardter, *Phys. Rev. B* **1993**, *47*, 11571.

4. Z.X. Tang, C.M. Sorensen, K.J. Klabunde, and G.C. Hadjipanayis, *Phys. Rev. Lett.* 1991, *67*, 3602.

5. C.B. Rong, D.R. Li, V. Nandwana, N. Poudyal, Y. Ding, Z.L.. Wang, H. Zeng and J.P. Liu, Advanced Materials, 18, 2984 (2006).

6. Y.P. Wang, Y. Li, C.B. Rong, and J. P. Liu, Nanotechnology, 18, 465701 (2007).

7. N. G. Akdogan, G. C. Hadjipanayis, and D. J. Sellmyer, , J. Appl. Phys., 105, 07A710 (2009).

8. M. Yue, Y.P. Wang, N. Poudyal, C.B. Rong, and J. P. Liu, J. Appl. Phys., 105, 07A708 (2009).

9. N. Poudyal, C.B. Rong and J. P. Liu, J. Appl. Phys., 107, in press (2010).

10. D. Givord, H. S. Li, and J. M. Moreau, Solid State Commun. **50**, 497 (1984)

11. R. Grossinger, X. C. Kou, R. Krewenka, et al., IEEE Trans. Magn. **26**, 1954 (1990).

12. J. F. Herbst, **63,** 819, Rev. Mod. Phys., (1991).

13. R. Skomski and J. M. D. Coey, *Permanent Magnetism*, Institute of Physics, Bristol 1999.

14. X.C. Kou, M. Dahlgren, R. Grossinger, and G. Wiesinger, J. Appl. Phys., 81, 4428 (1997).

Fig. 1 TEM images of (a) floating-at-top particles and (b) 3-h settle-down particles. (c) SEM image of slurry particles.

Fig. 2 SAXS curves of the floating-at-top and 3h settle-down particles. The inset gives the size distribution of the floating-at-top and 3h settle-down particles.

Fig. 3 (a) XRD patterns of the floating-at-top, 3h settle-down and slurry particles with epoxy protection. The XRD pattern of raw materials is also given for comparison. (b) XRD patterns of the 3h settle-down particles after exposure in air without epoxy protection.

Fig. 4 The temperature dependence of (a) magnetization, (b) dM/dT, (c) real component ($\chi'$) and (d) the imaginary component ($\chi''$) of AC susceptibility of the 3h settle-down particles, slurry and raw materials.

Table 1 $T_{sr}$ of the NdFeB nanoparticles measured by DC and AC methods

| Materials<br>Particle size | EEC ingots | | | MQP-C ribbons | | |
|---|---|---|---|---|---|---|
| | $T_{sr}$(DC) | $T_{sr}$(AC) | $T_{sr}$(Ave) | $T_{sr}$(DC) | $T_{sr}$(AC) | $T_{sr}$(Ave) |
| ~ 45 µm (raw) | 134±2 | 135±3 | 135±3 | 119±3 | 120±2 | 120±3 |
| 300 nm (slurry) | 113±5 | 122±5 | 118±5 | 109±5 | 117±4 | 113±5 |
| 20 nm (3h settle) | 105±5 | 100±5 | 103±5 | 90±5 | 85±5 | 88±5 |

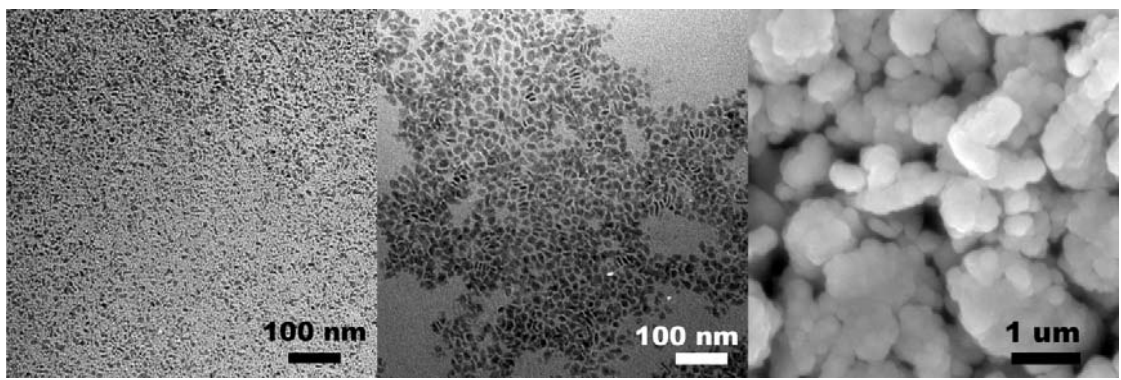

Fig. 1

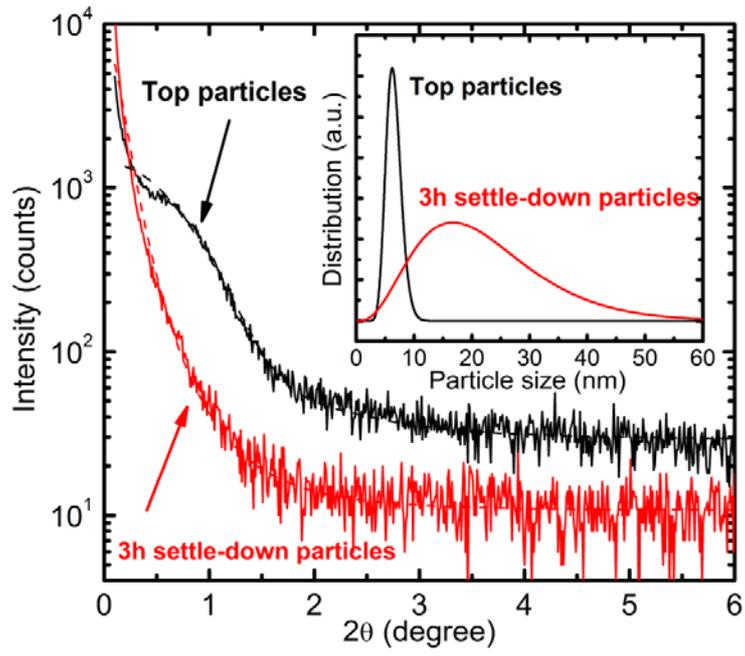

Fig. 2

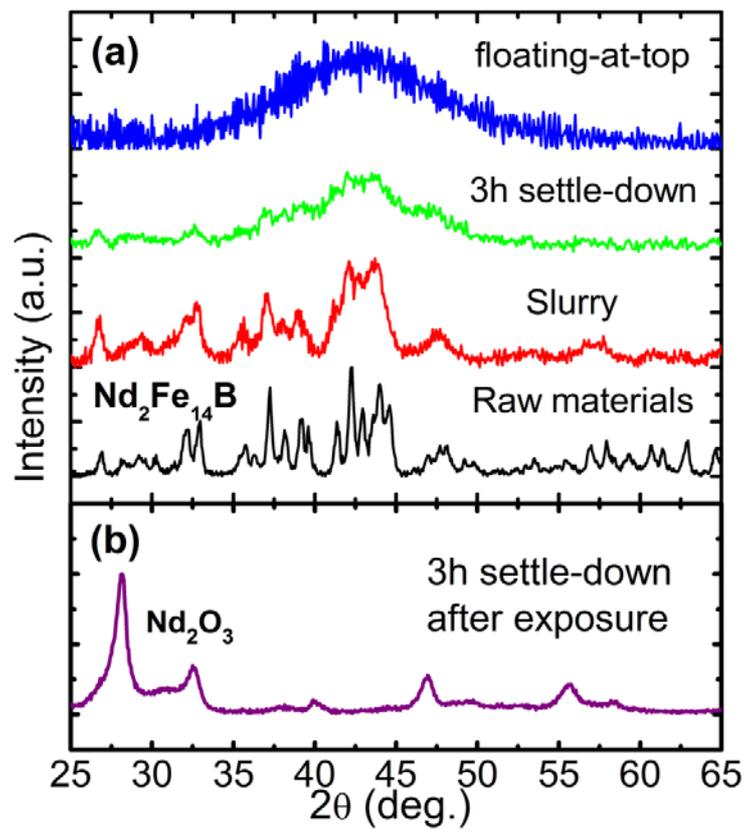

Fig. 3

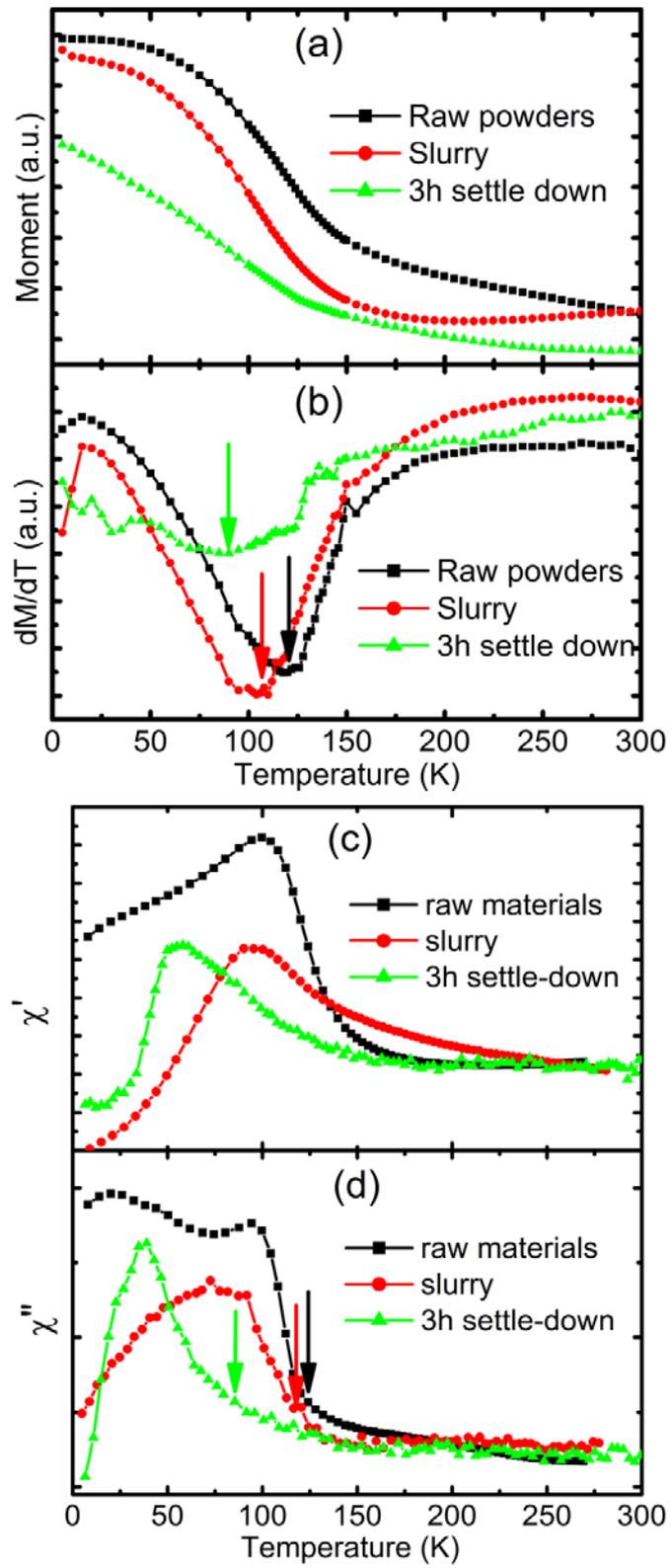

Fig. 4